\documentclass[preprint,showpacs,prb,amsmath]{revtex4}
\begin{document}

\title{Replacing Leads by Self-Energies\\
Using Nonequilibrium Green's Functions}

\author{Fredrick Michael and M.D. Johnson}
\email{mjohnson@ucf.edu}

\affiliation{Department of Physics, University of Central Florida, Orlando,
FL 32816-2385}

%\date{\today}
\date{March 22, 2002}

\begin{abstract}
An open quantum system consists of leads connected to a device of interest.
Within the nonequilibrium Green's function technique, we examine the
replacement of leads by self-energies in continuum calculations.
Our starting point is a formulation of the problem for continuum
systems by T.E.~Feuchtwang.  In this approach there is considerable
flexibility in the choice of unperturbed Green's functions. We examine
the consequences of this freedom on the treatment of leads.  For any
choice the leads can be replaced by coupling self-energies which are
simple functions of energy. We find that the retarded self-energy $\Sigma^r$
depends on the details of the choice of unperturbed Green's function,
and can take any value. However, the nonequilibrum self-energy
or scattering function $\Sigma^<$ can be taken to be independent
of this choice. Expressed in terms of these self-energies,
nonequilibrium transport calculations take a particularly simple form.
\end{abstract}

\pacs{73.23.-b, 73.63.-b,73.40.-c}

\maketitle

\section{Introduction\label{intro}}

A general method for calculating electronic transport properties 
is provided by the use of non-equilibrium Green's functions 
(NEGF).\cite{kadanoff,keldysh,langreth,rammer,haug} Because it has a
fully quantum mechanical and non-equilibrium formulation, and
because its foundational assumptions are transparent, this
technique has been used to study a very wide range of transport
problems.

In this paper we will examine the way in which the leads permitting
transport through an open quantum system (a ``device'')
can be replaced by self-energy
terms in a Dyson's equation. The analysis is limited to the case where the
leads can be well-approximated as systems of non-interacting electrons
in local thermodynamic equilibrium. In principle this approximation
is not very restrictive, since sufficiently far even from a mesoscopic
device the wires carrying current become macroscopic. Notice that
the device whose transport properties are being calculated is left
completely unspecified -- it can contain interactions, disorder, 
{\itshape etc}.

A similar treatment of the leads has been used by Datta, in an
approximation in which the spatial continuum
is replaced by a lattice of discrete points.\cite{datta}
An apparently very different approach for continuum systems was
developed in a series of papers by Caroli and co-workers\cite{caroli}
and by Feuchtwang.\cite{feuchtwang} The Feuchtwang approach
has been the basis of NEGF calculations by, for example,
Ref.~\onlinecite{schwabe}. Other workers have also cast the
continuum problem into a self-energy form.\cite{wang}

Our point here is to show that the Feuchtwang approach for the continuum
also can be repackaged so that leads are replaced exactly by simple 
self-energies, and to explore this possibility in some detail.
We make two simple changes to Feuchtwang's calculation: regrouping
the Dyson's equations to eliminate differential operators, and then
using analytic continuation as provided by Langreth's
theorem\cite{langreth} (and learned by us from 
Haug and Jaugho\cite{haug}; we will refer to this admirable
textbook for all standard information about NEGF).
The result is in our opinion pleasingly clear,
and simplifies transport calculations of electronic devices.

There is some freedom in how the self-energy is identified.
A Hamiltonian $H$ can in principle be arbitrarily
separated into an unperturbed $H_0$ and a
perturbation $H-H_0$. If calculations can then be
carried out exactly, the results will be independent
of the initial separation.  
There is a similar arbitrariness in the definition
of the unperturbed retarded $g^r$ and its corresponding
retarded self-energy $\Sigma^r$. (In the Feuchtwang approach
this comes from an arbitrary choice of internal boundary
conditions). The exact Green's functions are of course
independent of this choice.
It turns out, moreover, that the ``less-than'' self-energy
$\Sigma^{<}$ is independent of this choice, for a broad class
of internal boundary conditions. 

For clarity we will focus on the simplest case: a one-dimensional
system of non-interacting, spinless electrons in steady-state. But the results
are rather general. 
For example, electron-electron and other interactions can be included
by adding their appropriate self-energies to the coupling 
self-energy described here. 

Extension to two and three dimensions is also straightforward.
One motivation for this paper is that the approach we describe turns out
to work well even for systems with irregular shapes.\cite{mj}
Because the final results are independent of the choice of internal
boundary conditions, in practice one should make whatever choice
is simplest. In one dimension any choice is equally easy, but in
higher dimensions with irregular shapes, the 
right choice can simplify the calculation considerably.\cite{mj}

The model we will consider is an open one-dimensional system consisting
of two leads L and R (at $x<L$ and $x>R$, respectively), connected
to a device D (within $L<x<R$).  This has Hamiltonian\cite{feuchtwang}
\begin{equation}
H = \theta(L-x)H_L  + \theta(x-L) \theta(R-x)H_D  + \theta(x-R) H_R.
\label{H}
\end{equation}
where $\theta$ is a step function (0 to 1) and
\begin{equation}
H_{\alpha} = -\frac{\hbar^2}{2m}\frac{\partial^2}{\partial x^2} +
V_{\alpha}(x), \qquad
\alpha=L,D,R.
\label{Halpha}
\end{equation}
The single-particle potential energies $V_{\alpha}(x)$ are arbitrary.
To give definite examples in the following we will usually assume that
$V_{L,R}$ (but not $V_D$) are constants.
The leads are treated as reservoirs
of non-interacting electrons in local thermodynamic equilibrium described
by local chemical potentials $\mu_L,\mu_R$. 
An applied bias $eV$ would
be included by setting $\mu_L=\mu_R+eV$ and perhaps $V_L=V_R+eV$.
This would lead to a current
flow $I$, and a typical purpose of a NEGF calculation is to find the
I-V characteristics of the device. 
The current and other physical observables of a non-equilibrium system
are obtained from the
total ``less-than'' function $G^{<}$. We concentrate here on finding $G^{<}$.
The path we follow is to calculate the retarded Green's function $G^r$ 
and then use analytic continuation rules to obtain $G^<$.

\section{Retarded Green's Function $G^r$}

When the Hamiltonian is independent of time, the retarded Green's function
$G^r=G^r(x_1t_1,x_2t_2)$ can be Fourier
transformed in $t_1-t_2$. For brevity denote the result
$G(12)=G(x_1x_2\omega)$.
For a system of noninteracting electrons this satisfies
the equation of motion
\begin{equation}
[\hbar\omega - H(1) + i\eta] G^r(12) = \delta(x_1-x_2).
\label{eqnG}
\end{equation}
Here $H(1)$ is the total Hamiltonian in Eq.~(\ref{H}) written in
terms of coordinate
$x_1$, and $\eta>0$ is infinitesimal. The advanced Green's functions,
used below, are simply obtained by $G^a(12) = [G^r(12)]^*$.

\subsection{Unperturbed $g^r$}

Following Feuchtwang,\cite{feuchtwang}
one can similarly define individual ``unperturbed'' retarded Green's
functions associated with each region $\alpha=L,D,R$:
\begin{equation}
[\hbar\omega - H_{\alpha}(1) + i\eta] 
g^r_{\alpha}(12)
= \delta(x_1-x_2),
\label{eqng}
\end{equation}
subject to {\itshape internal} boundary conditions at $x=L,R$. Like Feuchtwang
we will consider the
set of possible homogeneous boundary conditions
\begin{subequations}
\label{bc}
\begin{eqnarray}
\left.\frac{\partial g_L^r(12)}{\partial x_1}\right|_{x_1=L}
&=& \gamma_L\, g_L^r(L2),\quad \mbox{if\ }x_2< L \\
\left.\frac{\partial g^r_D(12)}{\partial x_1}\right|_{x_1={L}}
&=& \gamma_{DL}\, g_D^r(L2),\quad \mbox{if\ }L< x_2< R \\
\left.\frac{\partial g^r_D(12)}{\partial x_1}\right|_{x_1={R}}
&=& \gamma_{DR}\, g_D^r(R2),\quad \mbox{if\ }L< x_2< R \\
\left.\frac{\partial g^r_R(12)}{\partial x_1}\right|_{x_1=R}
&=& \gamma_R\, g_R^r(R2),\quad \mbox{if\ }x_2> R,
\end{eqnarray}
\end{subequations}
plus identical conditions with $x_1$ and $x_2$ interchanged.
Unlike Feuchtwang, we do not assume that the $\gamma$ are all equal.
(This is important in higher dimensions.\cite{mj})
Eqs.~(\ref{bc}) include Dirichlet and von Neumann
boundary conditions as the limiting cases
$\gamma\rightarrow\infty,0$, respectively. The total Green's functions
$G$ cannot, and as we will see do not, depend on these internal
boundary conditions.

One also needs boundary conditions at the system's physical
boundaries $x_i\rightarrow\pm\infty$. 
These can be any homogeneous conditions.
As long as we choose $G$ to have the same homogeneous conditions on
the physical boundaries, the boundary conditions there play no role.

Both the perturbed and unperturbed retarded functions have matching
conditions of the form
\begin{equation} 
\label{match}
\left.\frac{\partial g^r(12)}{\partial x_1}\right|_{x_1=x_2-\epsilon}^{x_2+\epsilon}
=\frac{2m}{\hbar^2},\qquad
\left.g^r(12)\right|_{x_1=x_2-\epsilon}^{x_2+\epsilon}=0
\end{equation}
for infinitesimal $\epsilon>0$.
The first of these comes from integrating Eq.~(\ref{eqng}) across
an infinitesimal interval about $x_1=x_2$.

When $V_L$ and $V_R$ are constant, it is simple to obtain the unperturbed
$g_{\alpha}^r$ in the leads ($\alpha=L,R$).
Solve Eq.~(\ref{eqng})
separately in the regions $x_1<x_2$ and $x_1>x_2$, and then use
the boundary conditions Eq.~(\ref{bc}a,d) and the
matching conditions Eq.~(\ref{match}) to fix the undetermined constants.
The infinitesimal $\eta$ in Eq.~(\ref{eqng})
serves to pick out the outgoing (retarded) solution and
can then be set to zero.  The results are
\begin{subequations}
\label{gr}
\begin{eqnarray}
g_L^r(12) &=& \frac{2m}{\hbar^2 k_L} e^{-i[k_L(x_< -L)+\phi_L]}
\sin[{k_L(x_>-L) + \phi_L}]\\
g_R^r(12) &=& -\frac{2m}{\hbar^2 k_R} e^{i[k_R(x_> -R)+\phi_R]}
\sin[{k_R(x_<-R) + \phi_R}]
\end{eqnarray}
\end{subequations}
where $x_>$ ($x_<$) is the greater (lesser) of $x_1$ and $x_2$, and
where 
\begin{subequations}
\begin{eqnarray}
\hbar\omega &=& V_{\alpha} + \frac{\hbar^2k_{\alpha}^2}{2m},\label{kr}\\
\gamma_{\alpha} &=& k_{\alpha}\cot\phi_{\alpha},\label{phi}
\end{eqnarray}
\end{subequations}
for $\alpha=L,R$. If $\gamma_{L,R}$ are real, so too are $\phi_{R,L}$.
It is apparent in Eqs.~(\ref{gr}) that the unperturbed $g^r(12)$ are 
symmetric under $x_1\leftrightarrow x_2$. This is true for all
other Green's functions as well.

\subsection{Total $G^r$}

Once more following Feuchtwang,\cite{feuchtwang} 
consider the following integral over lead L:
\begin{eqnarray}
\int_{-\infty}^L d x_3 \left\{
G^r(32)[ \hbar\omega - H_L(3) + i\eta] g_L^r(13) -
g_L^r(13)[ \hbar\omega - H(3) + i\eta] G^r(32) \right\}\nonumber\\
 = \theta(L-x_1)G^r(12) - \theta(L-x_2)g_L^r(12) .
\label{integral}
\end{eqnarray}
The equality follows from the equations of motion, Eqs.~(\ref{eqnG},\ref{eqng}).
On the left-hand side of Eq.~(\ref{integral}), all the terms in the two square
brackets cancel except those involving derivatives. Hence the above integral
also equals
\begin{eqnarray}
\frac{\hbar^2}{2m} \int_{-\infty}^L d x_3 \left(
G^r \frac{\partial^2g_L^r}{\partial x_3^2} - g_L^r \frac{\partial^2 G^r}{\partial x_3^2} \right)\nonumber\\
= \frac{\hbar^2}{2m}
\left[ G^r(32)\frac{\partial g_L^r(13)}{\partial x_3}
- g_L^r(13)\frac{\partial G^r(32)}{\partial x_3}\right]_{x_3=L}.
\label{integral2}
\end{eqnarray}
Here the right-hand side follows from Green's theorem. 
(There are no contributions from the physical boundary $x_3\rightarrow-\infty$ 
as long as $g_L^r$ and $G$ have identical homogeneous conditions there.)
Equating the right-hand sides of
Eqs.~(\ref{integral},\ref{integral2}), yields
\begin{subequations}
\label{dyson}
\begin{equation}
G^r(12) = g_L^r(12)\theta(L-x_2)+\frac{\hbar^2}{2m}
\left[
 \frac{\partial g_L^r(13)}{\partial x_3} G^r(32)
- g^r_L(13)\frac{\partial G^r(32)}{\partial x_3}
\right]_{x_3=L}
(x_1 < L).
\end{equation}
We can perform similar integrals over the other two regions, replacing
$g_L^r$ by $g_D^r$ or $g_R^r$ as appropriate, with the result:
\begin{eqnarray}
G^r(12) &=& g_D^r(12)\theta(x_2-L)\theta(R-x_2)\nonumber\\
&+&\frac{\hbar^2}{2m}
\left[
\frac{\partial g_D^r(13)}{\partial x_3} G^r(32) 
- g_D^r(13) \frac{\partial G^r(32)}{\partial x_3} 
\right]_{x_3=L}^{R} (L< x_1 < R),\\
G^r(12) &=& g_R^r(12)\theta(x_2-R)-\frac{\hbar^2}{2m}
\left[
 \frac{\partial g_R^r(13)}{\partial x_3} G^r(32)
- g^r_R(13)\frac{\partial G^r(32)}{\partial x_3}
\right]_{x_3=R}
(x_1 > R).
\end{eqnarray}
\end{subequations}
Each of Eqs.~(11) is of the form of a Dyson equation ($G^r=g^r+g^r\Sigma^rG^r$).
They can be assembled into a single Dyson's equation for all 
space.\cite{feuchtwang}
However, the self-energies include differential operators, an
unattractive feature.

\subsection{Replace the leads by coupling self-energies\label{self}}

Eliminating the derivatives gives a simpler expression. 
This is where we diverge from Feuchtwang's path.
Let us focus on the only case of interest, $G^r(12)$ when both arguments
lie within the device ($L< x_1,x_2 < R$).
This is given by Eq.~(\ref{dyson}b). Replacing
$\partial g^r_D/\partial x_3$ at the internal
boundaries using Eq.~(\ref{bc}b,c), 
\begin{eqnarray}
G^r(12) = g_D^r(12) &+& \frac{\hbar^2}{2m} g_D^r(1R)
\left[ \gamma_{DR}G^r(R2) - 
\left.\frac{\partial G^r(32)}{\partial x_3}\right|_{x_3=R} 
\right] \nonumber\\
&-& \frac{\hbar^2}{2m} g_D^r(1L)
\left[ \gamma_{DL}G^r(L2) - 
\left.\frac{\partial G^r(32)}{\partial x_3}\right|_{x_3=L} 
\right] 
\label{dyson2}
\end{eqnarray}
when $L<x_1,x_2<R$.
Here we need $\partial G^r(32)/\partial x_3$ at
$x_3=L^+$.  This is obtained from Eq.~(\ref{dyson}a) by using 
Eq.~(\ref{bc}a), choosing $x_2>L$, and
letting $x_1\rightarrow L^-$:
\begin{equation}
\left.\frac{ \partial G^r(12)}{\partial x_1}\right|_{L^-} 
= \left( \gamma_L - \frac{2m}{\hbar^2}\frac{1}{g_L^r(LL)}\right)
G^r(L2) , \quad x_2> L.
\label{dGL}
\end{equation}
The total $G^r$ and its derivative are continuous 
across the internal boundaries.  Consequently
Eq.~(\ref{dGL}) (and a similar expression for the derivative at $R$) give
\begin{subequations}
\label{Grderiv}
\begin{eqnarray}
\left.\frac{ \partial G^r(32)}{\partial x_3}\right|_{L} 
&=& \left( \gamma_L - \frac{2m}{\hbar^2}\frac{1}{g_L^r(LL)}\right)
G^r(L2) , \quad x_2> L\\
\left.\frac{ \partial G^r(32)}{\partial x_3}\right|_{R} 
&=& \left( \gamma_R + \frac{2m}{\hbar^2}\frac{1}{g_R^r(RR)}\right)
G^r(R2), \quad x_2< R.
\end{eqnarray}
\end{subequations}
Plugging these derivatives into Eq.~(\ref{dyson2}) gives the following
Dyson equation for $G^r$ in the device:
\begin{eqnarray}
G^r(12) = g_D^r(12) + 
g_D^r(1L) \Sigma^r_L(\omega) G^r(L2) + 
       g_D^r(1R) \Sigma^r_R(\omega) G^r(R2)
\nonumber\\
(L < x_1,x_2 < R),
\label{Gr}
\end{eqnarray}
where
\begin{subequations}
\label{sigma}
\begin{eqnarray}
\Sigma^r_L(\omega) &=& -\frac{\hbar^2}{2m}
(\gamma_{DL}-\gamma_L) - \frac{1}{g_L^r(LL)},\\
\Sigma^r_R(\omega) &=& \phantom{-}\frac{\hbar^2}{2m}
(\gamma_{DR}-\gamma_R) - \frac{1}{g_R^r(RR)}.
\end{eqnarray}
\end{subequations}
Thus for the purpose of finding $G^r$ in the device, the
contribution of the leads has been replaced by simple
coupling self-energies $\Sigma^r_{L,R}$. Notice that these
are now $c$-numbers and not operators. This result is correct
even for Dirichlet conditions as long as they are obtained as
the limit $\gamma_{\alpha}\rightarrow\infty$.
One can get a final closed expression for $G^r(12)$ by evaluating
Eq.~(\ref{Gr}) at $x_1=L,R$, solving the resulting pair of equations
for $G^r(L2),G^r(R2)$, and substituting the results back into Eq.~(\ref{Gr}).

The standard case will have a constant potential $V_{L,R}$ in
the two leads, in which case we can substitute the particular
expressions Eqs.~(\ref{gr}) into Eqs.~(\ref{sigma}):
\begin{subequations}
\label{sigma2}
\begin{eqnarray}
\Sigma^r_L(\omega) &=& \frac{\hbar^2}{2m}(-\gamma_{DL} - 
i k_L)\\
\Sigma^r_R(\omega) &=& \frac{\hbar^2}{2m}(\phantom{-}\gamma_{DR} - 
i k_R).
\end{eqnarray}
\end{subequations}
We see that the choice of boundary condition in the leads drops
out completely --- there is no dependence on $\gamma_{L,R}$ left in
Eqs.~(\ref{sigma2}). However, the retarded self-energies $\Sigma^r$
do depend on the boundary conditions for the unperturbed $g^r_D$ in
the device, via $\gamma_{DL}$ and $\gamma_{DR}$. We emphasize again
the total $G^r$ solved from Eq.~(\ref{Gr}) will be independent
of this choice, also.

These results agree with the discretized formulation by Datta\cite{datta}
when the latter is taken to a continuum limit.
Datta discretizes the coordinates into lattice
points $ja$ (with $j$ an integer). In the Hamiltonian the kinetic energy
term becomes a second difference. The step from the last site in 
lead L (at $L-a$, say) and the first site in the device (at $L$) becomes
a perturbation. (A similar perturbation connects $R$ in the device to
$R+a$ in lead R.) Then an analysis like the above shows that 
the contribution of lead L to the
discrete $\tilde{G}^{r}$ in the device can be written in terms of a self-energy
\begin{equation}
\tilde{\Sigma}_L^{r} = -\frac{\hbar^2}{2ma^2}e^{i\tilde{k}_La}
\label{sigp}
\end{equation}
where
\begin{equation}
\hbar\omega = \frac{\hbar^2}{ma^2}(1-\cos{\tilde{k}_La}).
\end{equation}
For small lattice constant $a$, $\tilde{k}_L\approx k_L$ and Eq.~(\ref{sigp})
becomes
\begin{equation}
a\tilde{\Sigma}_L^{r} \approx \frac{\hbar^2}{2m} (-a^{-1} -ik_L).
\label{sigpp}
\end{equation}
In the discrete approximation the unperturbed $\tilde{g}_D^{r}$ in the device
vanishes at $L-a$. Thus for small $a$,
\begin{equation}
\tilde{g}_D^{r}(L-a,x_2) = 0 \approx \tilde{g}_D^{r}(Lx_2) - a
\left.\frac{\partial \tilde{g}_D^{r}(12)}{\partial x_1}\right|_{x_1=L}.
\end{equation}
That is, the discrete calculation for small lattice constant $a$ is
equivalent to homogeneous boundary conditions with $\gamma_{DL}=a^{-1}$.
Then the discrete self-energy Eq.~(\ref{sigpp}) agrees in the
continuum limit with our general result Eq.~(\ref{sigma2}a).

Because the boundary constants $\gamma_{DL},\gamma_{DR}$ are 
arbitrary, one can obviously choose $\Sigma_{L,R}^r$ with complete freedom.
Probably the simplest case is to have von Neumann boundary conditions
in the device ($\gamma_{DL}=0=\gamma_{DR}$). This choice is the only
simple one in higher dimensions for devices of varying cross section.\cite{mj}
But every other choice is also possible. One can even choose $\Sigma^r_{L,R}$
to vanish (in which case $\gamma_{DL}=-ik_L$ and $\gamma_{DR}=ik_R$ are
purely imaginary).
The latter case amounts to building the
contribution of the leads into the unperturbed $g^r_D$.

\section{Nonequilibrium Green's function $G^{<}$}

In symbolic notation, the Dyson's equation Eq.~(\ref{Gr}) for
$G^r$ in the device is
\begin{equation}
G^r = g_D^r + g_D^r\Sigma^r G^r.
\label{abbrev}
\end{equation}
Standard analytic continuation rules\cite{haug} then indicate
that the exact contour-ordered Green's function $G$ in the device
satisfies
\begin{equation}
G = g_D + g_D \Sigma G,
\label{G}
\end{equation}
where $g_{\alpha}$ is an unperturbed contour-ordered Green's function and
$\Sigma$ is the contour-ordered analogue of $\Sigma^r$.
Then a further application of the analytic continuation rules yields
\begin{equation}
G^< = g_D^< + g_D^r \Sigma^r G^< + g_D^r \Sigma^< G^a + g_D^< \Sigma^a G^a .
\label{Gless}
\end{equation}
Here $G^a(12)=[G^r(12)]^*$
is the advanced total Green's function and $\Sigma^a=(\Sigma^r)^*$ the
advanced self-energy. 

But what is the ``less-than" self-energy $\Sigma^{<}$? Our 
expressions Eqs.~(\ref{sigma}) for the retarded self-energies $\Sigma^r_{L,R}$
involve inverses, and can't be analytically continued using the standard
rules.\cite{haug}
One could make the following physical argument:\cite{datta} the self-energies
considered here represent the coupling to leads which are in local equilibrium;
hence one should be able to use the relationship between retarded
and less-than functions valid for equilibrium quantities. That is,
\begin{subequations}
\label{guess}
\begin{equation}
\Sigma^{<}_{\alpha}(\omega) = i f_{\alpha}(\omega) \Gamma_{\alpha}(\omega)
\end{equation}
where
\begin{equation}
\Gamma_{\alpha}(\omega) = i \left[ \Sigma^r_{\alpha}(\omega)-
\Sigma^a_{\alpha}(\omega)\right],
\end{equation}
\end{subequations}
for $\alpha=L,R$. Here
$f_{\alpha}(\omega)=1/(\exp[\beta(\hbar\omega-\mu_{\alpha})]+1)$
is the Fermi-Dirac equilibrium distribution for region $\alpha$.
As long as $\gamma_{DL},\gamma_{DR}$ are real, substituting 
Eqs.~(\ref{sigma2}) into Eqs.~(\ref{guess}) gives
\begin{equation}
\Sigma^{<}_{\alpha}(\omega) = if_{\alpha}(\omega)\frac{\hbar^2 k_{\alpha}}{m}
\theta(\hbar\omega-V_{\alpha}),
\quad \alpha=L,R.
\label{sigless}
\end{equation} 
While this physical argument may not be entirely convincing, it has
the merit of being correct (for real $\gamma_{DL},\gamma_{DR}$), as
we will now show.

\subsection{Unperturbed $g^<$}

Before coupling the three regions, each is in local equilibrium
with local chemical potential $\mu_{\alpha}$ ($\alpha=L,D,R$).
Consequently we can use the equilibrium relationship between
$g^<$ and $g^r$:
\begin{subequations}
\label{gless}
\begin{equation}
g_{\alpha}^<(12) = i f_{\alpha}(\omega) A_{\alpha}(12) \\
\end{equation}
where the spectral function in each region is:
\begin{equation}
A_{\alpha}(12) = i \left[ g^r_{\alpha}(12) - g^a_{\alpha}(12) \right].
\end{equation}
\end{subequations}
When all of the $\gamma$ in Eqs.~(\ref{bc}) are real, each $g^<_{\alpha}$
satisfies boundary conditions exactly like those in Eqs.~(\ref{bc}).
We will restrict ourselves to this case until the end of
this section.

When $V_{L,R}$ are constant, the unperturbed $g^<$ in the leads 
($\alpha=L,R$) are
\begin{equation}
g_{\alpha}^<(12) = i f_{\alpha}(\omega) 
\theta(\hbar\omega-V_{\alpha})
\frac{4m}{\hbar^2 k_{\alpha}} 
\sin[k_{\alpha} (x_1-\alpha) + \phi_{\alpha}] 
\sin[k_{\alpha} (x_2-\alpha) + \phi_{\alpha}],
\end{equation}
using Eqs.~(\ref{gr},\ref{gless}).
The unperturbed $g_D^<$ in the finite device is a sum over Dirac
$\delta$-functions. We will show later that it plays no role.

\subsection{Analytic continuation}

We calculate the total $G^<$ using the analytic continuation rules summarized
by\cite{haug}
\begin{subequations}
\label{ac}
\begin{eqnarray}
(AB)^r &\rightarrow& A^r B^r \\
(AB)^< &\rightarrow& A^r B^< + A^< B^a .
\end{eqnarray}
\end{subequations}
Here $A$ (and $B$) are contour-ordered quantities, and $A^r,A^<$ are
the retarded and less-than components of $A$.
Begin with Eqs.~(\ref{dyson}). Using Eq.~(\ref{ac}a), each of these 
can be analytically continued back to contour-ordered expressions.
(Simply remove the superscript $r$). Then
analytically continuing again with Eq.~(\ref{ac}b) we have
\begin{subequations}
\label{Glessac}
\begin{eqnarray}
G^<(12) &=& \frac{\hbar^2}{2m}
\left[
 \frac{\partial g_L^r(13)}{\partial x_3} G^<(32)
- g^r_L(13)\frac{\partial G^<(32)}{\partial x_3}\right. \nonumber\\
&& \qquad + \left.\frac{\partial g_L^<(13)}{\partial x_3} G^a(32)
- g^<_L(13)\frac{\partial G^a(32)}{\partial x_3}
\right]_{x_3=L}
\quad (x_1 < L, x_2 > L) ,\\
G^<(12) &=& g_D^<(12)
+\frac{\hbar^2}{2m}
\left[
\frac{\partial g_D^r(13)}{\partial x_3} G^<(32) 
- g_D^r(13) \frac{\partial G^<(32)}{\partial x_3}\right. \nonumber\\
&& \qquad + 
\left.\frac{\partial g_D^<(13)}{\partial x_3} G^a(32) 
- g_D^<(13) \frac{\partial G^a(32)}{\partial x_3} 
\right]_{x_3=L}^{R} \quad (L < x_1,x_2 < R),\\
G^<(12) &=& -\frac{\hbar^2}{2m}
\left[
 \frac{\partial g_R^r(13)}{\partial x_3} G^<(32)
- g^r_R(13)\frac{\partial G^<(32)}{\partial x_3}\right.\nonumber\\
&& \qquad + \left. \frac{\partial g_R^<(13)}{\partial x_3} G^a(32)
- g^<_R(13)\frac{\partial G^a(32)}{\partial x_3}
\right]_{x_3=R} \quad
(x_1 > R, x_2 < R).
\end{eqnarray}
\end{subequations}
The derivatives here can be eliminated much as they were in the
Dyson equation for $G^r$. Eqs.~(\ref{Glessac}a,c) yield the derivatives
$\partial G^</\partial x$ at $L,R$. The complex conjugate of
Eqs.~(\ref{Grderiv}) gives the derivatives $\partial G^a/\partial x$.
Substituting all of these into Eqs.~(\ref{Glessac}b) gives the
following expression for $G^<$ in the device:
\begin{eqnarray}
\label{Gless2}
G^<(12) &=& g_D^<(12) + g_D^r(1L)\Sigma_L^r G^<(L2) + 
g_D^r(1L)\Sigma_L^<G^a(L2) \nonumber\\
&+& g_D^<(1L) \Sigma_L^a G^a(L2) + (L\rightarrow R)\qquad (L< x_1,x_2< R),
\end{eqnarray}
where $\Sigma^r_{L,R},\Sigma^a_{L,R}$ are given by Eqs.~(\ref{sigma}) 
and their complex conjugate, and
\begin{equation}
\label{sigless2}
\Sigma_{\alpha}^<(\omega) = \frac{g_{\alpha}^<(\alpha\alpha)}
{g_{\alpha}^r(\alpha\alpha) g_{\alpha}^a(\alpha\alpha)} \quad (\alpha=L,R).
\end{equation}
In the usual case the potential in the
leads $V_{L,R}$ is constant. Then, using Eqs.~(\ref{gr},\ref{gless}),
this becomes equal to Eq.~(\ref{sigless}).
Thus the physical argument used at the beginning of this section is
correct, at least as long as the $\gamma$'s are real.

Eq.~(\ref{Gless2}) is of the standard form for the analytic continuation of
a Dyson's equation for $G^r$ to an equation for $G^<$. It is easy to show
that Eq.~(\ref{Gless2}) can be written equivalently as
the Keldysh quantum kinetic equation\cite{haug}
\begin{equation}
G^{<} = (1+G^r\Sigma^r)g_D^{<}(1+\Sigma^a G^a) + G^r\Sigma^{<}G^a.
\label{qke}
\end{equation}

For a finite length device, the first term in Eq.~(\ref{qke}) 
vanishes.\cite{datta,schwabe,combescot}
There are various ways to see this. It is clear on physical grounds that
the exact Green's function including coupling to semi-infinite leads
cannot depend on the initial occupancy of the finite device. 
We could choose $\mu_D\rightarrow-\infty$,
in which case the device is initially unoccupied and the unperturbed
$g_D^<$ vanishes. Proof that
the term vanishes is based on the fact that the unperturbed spectral function
$A_D$ for a finite device is a sum of Dirac $\delta$-functions. The first
term written symbolically in Eq.~(\ref{qke}) is, in detail,
\begin{eqnarray}
\label{term}
g_D^<(12) &+& G^r(1L)\Sigma_L^r g_D^<(L2) + G^r(1R)\Sigma_R^r g_D^<(R2)\nonumber\\
&+& g_D^<(1L)\Sigma_L^a G^a(L2)+ g_D^<(1R)\Sigma_R^a G^a(R2)\nonumber\\
&+& G^r(1L)\Sigma^r_L g_D^<(LL) \Sigma_L^a G^a(L2) \nonumber\\
&+& G^r(1L)\Sigma^r_L g_D^<(LR) \Sigma_R^a G^a(R2) \nonumber\\
&+& G^r(1R)\Sigma^r_R g_D^<(RR) \Sigma_R^a G^a(R2) \nonumber\\
&+& G^r(1R)\Sigma^r_R g_D^<(RL) \Sigma_L^a G^a(L2) .
\end{eqnarray}
For the finite device we expand the unperturbed $g_D$'s in discrete energy
eigenstates $\psi_i(x)$ of $H_D$ with boundary conditions
like Eq.~(\ref{bc}b,c):
\begin{eqnarray} 
g_D^r(12) &=& \sum_{i} \frac{\psi_i(x_1)\psi_i(x_2)}
{\hbar\omega-\varepsilon_i+i\eta} \\
g_D^<(12) &=& -2\pi f_D(\omega) \sum_{i}
 \psi_i(x_1)\psi_i(x_2)\delta(\hbar\omega-\varepsilon_i) .
\end{eqnarray}
Since each term in Eq.~(\ref{term}) contains a factor of $g_D^<$, the
term can be nonzero only if $\hbar\omega$ equals some eigenenergy
$\varepsilon_i$. But in this case [and solving for $G^{r,a}(L2),G^{r,a}(R2)$
from Eq.~(\ref{Gr})], some algebra shows that the terms in Eq.~(\ref{term})
cancel exactly.

Consequently the total NEGF is given by the second term indicated
symbolically in Eq.~(\ref{qke}). In detail for us this is
\begin{equation}
G^<(12) = G^r(1L)\Sigma^<_LG^a(L2) + G^r(1R)\Sigma^<_RG^a(R2).
\label{qke2}
\end{equation}
It seems from this last equation that the scattering
functions $\Sigma^<_{L,R}$ must be independent of the original
choice of internal boundary conditions. After all, $G^r$, $G^a$,
and $G^<$ are independent of this choice, and these three
and $\Sigma^<$ are related by Eq.~(\ref{qke2}). However, this
conclusion is not quite correct. We have shown that the $\Sigma^<_{L,R}$ are
independent of the internal boundary conditions, and given by
\begin{equation}
\Sigma^{<}_{\alpha}(\omega) = if_{\alpha}(\omega)\frac{\hbar^2 k_{\alpha}}{m}
\theta(\hbar\omega-V_{\alpha}),
\quad \alpha=L,R,
\label{guessagain}
\end{equation} 
as long as the $\gamma$'s are all real.
However, when any of the $\gamma$ are complex, a similar
analysis shows that $\Sigma^<_{L,R}$ pick up extra terms; but 
the extra terms give a
cancelling contribution to Eq.~(\ref{qke2}). The accurate statement
is rather that one can always choose the scattering functions to be
of the form given by Eq.~(\ref{guessagain}).

\section{Summary}

In this paper we have analyzed the replacement of leads by coupling
self-energies in calculations of nonequilibrium Green's functions. 
This was based on a modification
of Feuchtwang's approach for continuum systems. In general
the form of a self-energy depends on the way in which the unperturbed
Green's functions are defined. The final Green's functions of course do not.
Here we investigated a class of homogeneous
boundary conditions at the junctions between device and leads.
Because these internal boundary conditions are completely arbitrary,
the retarded self-energy $\Sigma^r$ can take any value.
The ``less-than'' self-energy or scattering function
$\Sigma^<$, on the other hand, can be taken to be
entirely independent of the choice of boundary
conditions.

In one dimension no particular choice of boundary conditions
is easier than another.
In more complicated calculations, for example of two- and three-dimensional
devices with nonuniform cross-sections, the ideas to take away
from this work are: (1) The leads can be replaced by self-energies;
(2) All internal boundary conditions give the same final result;
so (3) Choose the boundary conditions which make the calculation easy.
Following this route in our opinion simplifies transport calculations in general
electronic devices of arbitrary cross-section.
We will explore this in a subsequent paper.\cite{mj}

%\begin{acknowledgements}
\acknowledgements
We acknowledge support from the NSF through grant DMR99-72683.
%\end{acknowledgements}

\end{document}